\documentclass[12pt]{article}
\usepackage{hyperref}

\begin{document}


\begin{center}
{\Large \bf  Compatibility of  the Dimensional Reduction and Variation Procedures for a Quadratic Curvature Model with a Kaluza-Klein Ansatz}

\vspace{7mm}

S. Ba{\c s}kal
\footnote{electronic address:
baskal@newton.physics.metu.edu.tr}
and S. {\c C}elik
\footnote{electronic address: sinancel@metu.edu.tr}

\vspace{3mm}

Physics Department, Middle East Technical University\\
06531, Ankara, Turkey
\end{center}

\vspace{6mm}

\begin{abstract}
	The introduction of extra dimensions is an invaluable strategy for the unification of gravity with other physical fields.  Nevertheless, the matter in hand is to be eventually reduced to the actual 4D spacetime.  The Kaluza-Klein theory is no exception to this well-known scheme. There are two procedures to obtain the field equations from a higher dimensional action.  One can either take the variation of the effective action in that higher dimension and then reduce the resulting equations or reduce the higher dimensional action to the actual 4D and henceforward take the variations with respect to the constituent fields of the theory.  Here, for the case of a quadratic curvature model with a Kaluza-Klein ansatz the field equations are obtained from the reduced action and compatibility of these two procedures is discussed in detail.\\
	\end{abstract}

\section{Introduction}
In 1919, Weyl introduced a scale invariant effective action with a quadratic curvature \cite{weyl1919}
\begin{equation}\label{wya}
\int d^{4}x  \sqrt{-g} \, R_{jkmn}R^{jkmn}
\end{equation}
as an alternative model to Einstein's gravity.  In 1974, Yang introduced the underlying gauge structure of gravity, where the Riemann tensor represents gravitational  field \cite{yang1974}.  In this work we shall refer  (\ref{wya}) as the Weyl-Yang (WY) action.
Since then and in between, numerous authors, to name a few: Stelle\cite{stelle1977}, Stephenson\cite{stephenson1958}, Kilmister and Newman\cite{kilmister1961}, Higgs\cite{higgs1959},  Fairchild and Edward \cite{fairchild1976} have favored this particular model in their articles.  In addition to being a scale invariant gauge model, it includes the solutions of Einstein`s equations as a first integral, while its non-desirable properties can mostly be circumvented \cite{fairchild1976,pavelle1975,pavelle1976}.  

The Kaluza-Klein (KK) theory unifies electromagnetism with gravitation.  Unifying
gravity with electromagnetism by some suitable trivial and/or non-trivial couplings with the generic $F^{n}, (DF)^{n}$ and $RF^{2}$ terms  in the action is shown to be interesting \cite{drummond1980,dereli2011,dereli2020}.  They account for  photon graviton oscillations \cite{raffelt1988} and also drew attention from an astrophysical point of view \cite{lambiase2004}.  However, including terms involving the Riemann, Ricci and curvature scalar contracted with the electromagnetic field $F$, seems to be done (apart from the well-known basic principles) rather freely, while the explicit form of the couplings is predetermined in the KK theory as they  arise uniquely from the reduction mechanism. 

Quadratic curvature Lagrangians are also considered in the framework of the KK theory\cite{gokturk1990,huang1988}.  Later on,  Baskal and Kuyrukcu gave the field equations and the
energy-momentum tensor for the WY type of KK gravity \cite{baskal2013}.  In that article they took the variation of the effective action in five dimensions by using the Palatini approach which considers the 5D connection and the 5D metric as independent variables and then they reduced the resulting equations to the actual four dimensions.  
 
This work can be considered as an extension to \cite{baskal2013}.  Here,  we shall reverse the succession of mechanisms involved therein.  We shall first reduce the 
5D WY action to four dimensions and thereafter we shall take the variations with respect to the constituent fields of the theory, namely with respect to the 4D connection, metric, electromagnetic vector potential and the scalar dilaton field, as they become manifest  after the reduction.   The main purpose of this work is to examine the compatibility of these two procedures where  the order of the implementations of variation and reduction mechanisms are reversed, within the context of  the WY model with a KK ansatz.

\section{Preliminaries}
The Kaluza-Klein theory operates on a  five dimensional manifold with a coordinate system 
\begin{equation}
	\hat{x}^{A}=(x^{a}, x^{5}) 
\end{equation}
where one of the spacelike dimensions is spontaneously compactified to a circle, whose radius is in the order of  Planck's length. 
The indices with capital Latin letters take values  $A=0,1,2,3,5$, whereas all the indices with lowercase Latin letters take values $a=0,1,2,3$.  
The  KK metric in five dimensions can conveniently be written as
	\begin{equation}
	\hat{G}(x^{j},x^{5})=G(x^{j})+\varphi(x^{j})[A(x^{j})+dx^{5}]\otimes\varphi(x^{j})[A(x^{j})+dx^{5}]
	\end{equation}
	where
	\begin{equation}
		G(x^{j})=\eta_{ab} \, {e}^{a}\otimes {e}^{b}
	\end{equation}
The following choice of  the orthogonal basis 1-forms 
	\begin{equation}
		\hat{E}^{a}(x^{j},x^{5}) = {e}^{a}( x^{j})\qquad \mbox{and} \qquad \hat{E}^{5}=\varphi(x^{j})(A(x^{j})+dx^{5})
	\end{equation}
 allows the KK metric  to  be expressed succinctly
	\begin{equation}\label{metricsuccint}
		\hat{G}(x^{j},x^{5})=\eta_{AB}\hat{E}^{A}\otimes\hat{E}^{B}
	\end{equation}
	with $\eta_{AB}=diag(-1,1,1,1,1)$ and  $\hat{E}^{A}=(\hat{E}^{a},\hat{E}^{5})$,  satisfying $\iota_{X_{B}}(\hat{E}^{A})=\delta^{A}_{B}$ .  
	In terms of  the coordinate basis,  
	f{\"u}nf-beins can be expressed as  $\hat{E}^{A}=h^{A}\,_{\mu}dx^{\mu} $, 
	where  $\mu =0,1,2,3,5$ .  All physical fields are independent of the fifth coordinate $x^{5}$.
	
	Taking  the metric (\ref{metricsuccint}) into account, the components of the Riemann tensor had been found earlier in \cite{gokturk1990} and in \cite{baskal2013}:
\begin{equation}\label{riemm1}
	\begin{array}{ll}
&	\hat{R}_{abmn}=R_{abmn}-\frac{1}{4}\varphi^{2}(2F_{ab}F_{mn}+F_{am}F_{bn}-F_{an}F_{bm})\\[3mm]
&	\hat{R}_{ab5m}=\frac{1}{2}\varphi D_{m}F_{ab}+\frac{1}{2}(2\varphi_{m}F_{ab}+\varphi_{b}F_{am}-\varphi_{a}F_{bm})\\[3mm]
&\hat{R}_{5a5m}=-\varphi^{-1}D_{m}\varphi_{a}-\frac{1}{4}\varphi^{2}F_{aj}F^{j}\,_{m}
\end{array}
\end{equation}

\section{Kaluza-Klein reduction for a quadratic curvature}

One way to obtain the field equations in the standard KK theory is to vary the 5D action
\begin{equation}\label{Ein5}
\int  \hat{R} \, \sqrt{-\hat {g}}  \, d^{5}x 
\end{equation}
with respect to the 5D metric $\hat{g}_{AB}$ which yields
\begin{equation}\label{G}
	\hat{G}_{AB}=0
\end{equation}	
where 
\begin{equation}\label{GAB}
	\hat{G}_{AB}\equiv\hat{R}_{AB}- \frac{1}{2}\hat{g}_{AB}\, \hat{R}
\end{equation}	
is the Einstein tensor in 5D.
Then  (\ref{G})  is reduced to 4D to  obtain equations of gravity  coupled with electromagnetism and  a scalar field.  One may well start from reducing (\ref{Ein5}) to 4D,
\begin{equation}\label{EHL}
	\int\varphi[R-\frac{1}{4}\varphi^{2}F^{ab}F_{ab}
	-2\varphi^{-1}D_{a}\varphi^{a}]\,\sqrt{-g} \,d^{4}x
\end{equation}
and then take the variations with respect to the emerging fields, namely $g_{ab}, A_{k},$ and $\varphi$.  The question is whether these two approaches produce the same results.  Equations obtained from those two procedures are shown to be equivalent with some straightforward manipulations \cite{celik2021}. 

\subsection{The 5D Palatini approach to the WY  action}
A similar line of thought  begs the same type of question for the WY  action in 5D. 
The five dimensional gravitational action
\begin{equation}
	{\cal I}=\frac{1}{16\pi\hat {G}}\int d^{5}x\hat{\mathcal{L}}
\end{equation}
explicitly  reads as
\begin{equation} \label{quadact5}
	{\cal I}=\frac{1}{16\pi\hat {G}}\int d^{5}x\sqrt{-\hat{g}}\hat{R}_{JKMN}\hat{R}^{JKMN}
\end{equation}
when the Lagrangian is formed from  a  5D quadratic curvature.\\

Using the Palatini method where  
variation of   (\ref{quadact5})  is taken directly with respect to the 5D connection $\hat{\Gamma}^{J}_{MN}$ and the 5D metric $\hat{g}^{AB}$ renders  the field equations :
\begin{equation}\label{WYeq}
 D_{K}\hat R^{K}\,_{JMN}=0
\end{equation}
and  the energy-momentum tensor
\begin{equation}\label{TABb}
	\hat{T}_{AB}=\hat{R}_{AKMN}\hat{R}_{B}\,^{KMN}
	-\frac{1}{4}\hat{g}_{AB}R_{JKMN}R^{JKMN}
\end{equation}
respectively, where
\begin{equation}
	\hat{T}_{AB}=\frac{1}{2\sqrt{-\hat{g}}}\frac{\delta \hat{\mathcal{L}}}{\delta \hat{g}^{AB}}
\end{equation}
Then one proceeds to reduce (\ref{WYeq}) and (\ref{TABb}) to obtain the field equations and the energy-momentum tensor  in the actual 4D spacetime \cite{baskal2013}.  In the next section, we shall reverse the order of this mechanism.  We shall first reduce (\ref{quadact5}) into 4D and then we shall take the variations with respect to the  fields emerging after the reduction.

\subsection{Reduced form of  the WY action}
	
We start by expressing the 5D  invariant by  a natural splitting the Riemann tensor in terms of its  spacetime, fifth and the mixed components,  also keeping  in mind that
$ \sqrt{-\hat{g}}=\sqrt{-g}\varphi $.
        Then, the action  becomes
\begin{equation}
	{\cal I}=\frac{1}{16\pi G}\int d^{4}x\sqrt{-g}\varphi(\hat{R}_{jkmn}\hat{R}^{jkmn}+4\hat{R}_{5kmn}\hat{R}^{5kmn}+4\hat{R}_{j5m5}\hat{R}^{j5m5})
\end{equation}
We define $L= \varphi L_{q}$, where
\begin{equation}
	L_{q}=\hat{R}_{jkmn}\hat{R}^{jkmn}+4\hat{R}_{5kmn}\hat{R}^{5kmn}+4\hat{R}_{j5m5}\hat{R}^{j5m5}
\end{equation}
Taking (\ref{riemm1})  into account  our action becomes
\begin{equation}\label{actionallterms}
	\begin{array}{ll}
		{\cal I}&=\frac{1}{16\pi G}{\displaystyle \int} d^{4}x\sqrt{-g}\,\bigg\{\varphi R_{jkmn}R^{jkmn}-\frac{3}{2}\varphi^{3}R^{j}\,_{kmn}F_{j}\,^{k}F^{mn} \\ [4mm]
		&+\frac{1}{8}\varphi^{5}[3F_{jk}F^{jk}F_{mn}F^{mn}+5(F_{jm}F^{mn}F_{ni}F^{ij})]+\varphi^{3}D_{k}F_{mn}D^{k}F^{mn} \\ [4mm]
		&+4\varphi^{-1}D_{m}\varphi_{n}D^{m}\varphi^{n}-2\varphi^{2} F^{mk}F^{n}\,_{k}D_{m}(D_{n}\varphi) \\ [3mm]
		&+4\varphi^{2}(D_{k}F_{mn}\varphi^{m}F^{kn}+D_{k}F_{mn}\varphi^{k}F^{mn})
		\\[3mm]
		&+6\varphi(\varphi_{k}\varphi^{k}F^{mn}F_{mn}+\varphi_{m}\varphi_{n}F^{mk}F^{n}\,_{k}) \bigg\}
	\end{array}
\end{equation}

\section{Field equations from the reduced action}
In this section we shall treat each term in (\ref{actionallterms}) one by  one and collect them at the end.  So we define
\begin{equation}
	{\cal I}_{i}=\int d^{4} x \, \mathcal{L}_{i}=\int d^{4}x \, \sqrt{-g} \,L_{i}
\end{equation}
where the subscript $i$ is representing each term.  This way we have the advantage of  analysing their equivalent forms, in addition to referring to the literature for some physical content.
\\

\par
$\bullet$   ${\cal I}_{1}$  \\
The action for the first term is
\begin{equation}
	{\cal I}_{1}=\int d^{4}x  \sqrt{-g} \, \varphi R_{jkmn}R^{jkmn}
\end{equation}

By treating the gravitational field $R_{jkmn}$ as a gauge field \cite{yang1974},  dynamics  resulting from a quadratic curvature Lagrangian has been studied extensively \cite{fairchild1976,pavelle1975,baskal1999} (and the references therein).  An immediate generalization would be the construction of  this formalism for the KK ansatz \cite{gokturk1990, baskal2013}.
In fact, it is these latter developments that paved the way for this current work.\\

Now, consider
\begin{equation}
	\delta\mathcal{L}_{1}=2\varphi R_{jkmn}\delta R^{jkmn}\sqrt{-g}
\end{equation}
to vary with respect to the connection.
Let us take the Palatini equation into account \cite{dinverno1992}:
\begin{equation}\label{palatiniequation}
	\delta R^{j}\,_{kmn}=D_{m}(\delta \Gamma^{j}_{kn})-D_{n}(\delta \Gamma^{j}_{km})
\end{equation}
and 
\begin{equation} \label{412}
	D_{m}(\sqrt{-g}\varphi R_{j}\,^{kmn}\delta \Gamma^{j}_{kn})=D_{m}(\sqrt{-g}\varphi R_{j}\,^{kmn})\delta \Gamma^{j}_{kn}+\sqrt{-g}\varphi R_{j}\,^{kmn}D_{m}\delta \Gamma^{j}_{kn}
\end{equation}
Substituting  (\ref{palatiniequation}) 
and using (\ref{412}) we have
where the first and third terms are tensor densities of weight +1.  Apart from a  total divergence, we have
\begin{equation}
	\delta{\cal I}_{1}=4\int d^{4}xD_{n}(\sqrt{-g}\varphi R_{j}\,^{kmn})\delta \Gamma^{j}_{km}
\end{equation}
The expression resulting from above is
\begin{equation}
	\gamma_{1}:=-4\sqrt{-g}D_{m}(\varphi R_{j}\,^{kmn})
\end{equation}
\par
Variation of $	{\cal I}_{1}$ with respect to $\varphi$ is
\begin{equation}
	\delta{\cal I}_{1}=\int d^{4}x\sqrt{-g}\delta\varphi R_{jkmn}R^{jkmn}
\end{equation}
So, we get
\begin{equation}
	\Phi_{1}:=\sqrt{-g}R_{jkmn}R^{jkmn}
\end{equation}
\par
In order to vary with respect to $g^{ab}$, we rewrite our Lagrangian in terms of the hidden metrics
Using the symmetry properties of the Riemann tensor and simplifying we have
\begin{equation}
	\delta\mathcal{L}_{1}=\varphi(2 R_{anjk}R_{b}\,^{njk}-\frac{1}{2}g_{ab} R_{mnjk}R^{mnjk})\sqrt{-g}\delta g^{ab}
\end{equation}
\par
We define the energy-momentum tensor $T_{ab}$ as the coefficient of $\delta g^{ab}$ such that
\begin{equation}\label{emtensor}
	T_{ab}\delta g^{ab}=\frac{1}{2}\frac{1}{\sqrt{-g}}\delta\mathcal{L}
\end{equation}
Therefore, we obtain
\begin{equation}
	T_{ab}=\varphi (R_{anjk}R_{b}\,^{njk}-\frac{1}{4}g_{ab} R_{mnjk}R^{mnjk})
\end{equation}

\par
$\bullet$  ${\cal I}_{2}$  \\
For the second term we have  
\begin{equation}\label{rf2}
	{\cal I}_{2}=(-3/2)\int d^{4}x\sqrt{-g}\varphi^{3} R^{j}\,_{kmn}F_{j}\,^{k}F^{mn}
\end{equation}

Early works dealing with Lagrangians given in the  above form are mostly formal mathematical studies in essence \cite{prasanna1973, horndeski1976, buchdahl1979}.   Later on the motivations for the inclusion of $RF^{2}$ became diverse\cite{drummond1980}. An exhausted list of  $R_{x}F^{2}$ non-minimal couplings are presented in \cite{balakin2005}, where $R_{x}$ refers to an element  of  the set $ R_{x}=\{R, R_{ij}, R_{ijnm}\}$,  along  with the condition that couplings  of  $F^{2}$  with respect to $R_{x}$ are linear.  These authors argue  that such couplings depict drastic conditions where the gravitational and electromagnetic fields are intense and  the speed of the gravitational-electromagnetic wave 
differs from that of  light in vacuum.	We see that one such particular coupling  as in (\ref{rf2})  singles out naturally through the reduction procedure of the model we have, and excludes all other forms that might be added arbitrarily by hand.

We first start by varying (\ref{rf2}) with respect to  $A_{i}$
\begin{equation}
	\delta {\cal I}_{2}=(-3/2) \int d^{4}x \, \sqrt{-g}\varphi^{3} \,  \{ R^{jk}\,_{mn}(D_{j}\delta  A_{k}-D_{k}\delta  A_{j})F^{mn}+R_{jk}\,^{mn}F^{jk}(D_{m}\delta  A_{n}-D_{n}\delta  A_{m}) \}
\end{equation}

Apart from total divergences, we obtain 
\begin{equation}
	\delta {\cal I}_{2}=6\int d^{4}x 	 \sqrt{-g}D_{j}(\varphi^{3} R^{ji}\,_{mn}F^{mn})\delta  A_{i}
\end{equation}
which gives the expression
\begin{equation}
	a_{2}:=6\sqrt{-g}D_{k}(\varphi^{3}R^{kj}\,_{mn}F^{mn})
\end{equation}

Now, we vary with respect to the connection $\Gamma^{j}_{nk}$ 
\begin{equation}
	\delta  {\cal I}_{2}=(-3/2){\displaystyle \int } d^{4}x\sqrt{-g}\varphi^{3}(\delta  R^{j}\,_{kmn})F_{j}\,^{k}F^{mn} 
	\end{equation}
to obtain 
\begin{equation}
	\delta  {\cal I}_{2}=3 \, \int d^{4}xD_{m}(\sqrt{-g}\varphi^{3}F_{j}\,^{k}F^{mn})\, \delta  \Gamma^{j}_{nk}
\end{equation}
So, we have
\begin{equation}
	\gamma_{2}:=3 \, D_{m}(\sqrt{-g}\varphi^{3}F_{j}\,^{k}F^{mn})
\end{equation}

\par
By varying with respect to $\varphi$
\begin{equation}
	\delta{\cal I}_{2}=(-9/2) \int d^{4}x\sqrt{-g} R^{j}\,_{kmn}F_{j}\,^{k}F^{mn}\varphi^{2}\delta\varphi
\end{equation}
we end up with 
\begin{equation}
	\Phi_{2}:=(-9/2) \sqrt{-g}\varphi^{2}R^{j}\,_{kmn}F_{j}\,^{k}F^{mn}
\end{equation}

\par
Varying with respect to the metric $g^{ab}$ yields
\begin{equation}
	\begin{array}{ll}
		\delta  {\cal I}_{2}=(-3/2) {\displaystyle \int} d^{4}x \,  \varphi^{3}(2R_{a kmn}F_{b}\,^{k}F^{mn} \\[4mm]
		+2R_{akmn}F_{b}\,^{k}F^{mn}
		+\frac{1}{2}g_{ab}R_{jkmn}F^{jk}F^{mn})\sqrt{-g}\delta  g^{ab}
	\end{array}
\end{equation}
Thus, we obtain
\begin{equation}
	T_{ab}=(-3/2) \varphi^{3}(R_{anjk}F_{b}\,^{n}+R_{bnjk}F_{a}\,^{n}-\frac{1}{4}g_{ab}R_{jkmn}F^{mn}) \, F^{jk}
\end{equation}

\par

$\bullet$    ${\cal  I}_{3}$ \\
The action for the third term is
\begin{equation}
	{\cal I}_{3}=(3/8)\int d^{4}x\sqrt{-g}\, \varphi^{5}F_{jk}F^{jk}F_{mn}F^{mn}
\end{equation}
From a mathematical point of view $F^{n}$ invariants were considered to be worthwhile  in their own right, \cite{escobar2014}.  The most interesting results arising from $F^{4}$ terms has been found the context of QED  in Minkowskian space,  when photons interact with a strong magnetic field  \cite{adler1971}.

Specifically, one can evaluate following two quadratic invariants  in terms of  the electric $E_{i}$ and magnetic $B_{i}$ field vectors 
\begin{equation}
	\begin{array}{ll}
		& F_{jk}F^{jk}F_{mn}F^{mn}= 4 (E^{2}-B^{2})^{2}  \\[3mm]
		& F^{jk}F_{km}F^{mn}F_{nj}= 2 (E^{2}-B^{2})^{2}+ 4 (E  \cdot B )^{2}
	\end{array}
\end{equation}
to grasp the strength of the fields as compared to the two fundamental quadratic invariants 
$F_{jk}F^{jk}= 2 (E^{2}-B^{2})$ and $F_{jk}\, ^{\ast } F^{jk}=(E \cdot B)$. \\

Now, we take the variation of ${\cal I}_{3}$ with respect to $A_{n}$.
Apart from total divergences we obtain 
\begin{equation}
	\delta {\cal I}_{3}=3 \int d^{4}x\, \big\{
	-D_{m}(\sqrt{-g}\varphi^{5}F^{ij}F_{ij}F^{mn})\, \delta A_{n}\big\}
\end{equation}
which yields  to  the following expression
\begin{equation}
	a_{3}:=-3 D_{k}(\sqrt{-g}\varphi^{5}F_{mn}F^{mn} F^{kj})
\end{equation}

\par
Variation of ${\cal I}_{3}$ with respect to  $\varphi$
gives
\begin{equation}
	{\delta\cal I}_{3}=(15/8)\int d^{4}x \sqrt{-g}\, \varphi^{4}F_{jk}F^{jk}F_{mn}F^{mn}\,\delta \varphi
\end{equation}
From above, we have
\begin{equation}
	\Phi_{3}:=(15/8) \sqrt{-g}\varphi^{4}F_{jk}F^{jk}F_{mn}F^{mn}
\end{equation}

\par
The variation of $\mathcal{L}_{3}$ with respect to $g^{ab}$ becomes
\begin{equation}
	\delta\mathcal{L}_{3}=(3/8) \varphi^{5} \big\{F_{mn}F^{mn} \,(F_{aj}F_{b}\,^{j}
	+F^{j}\,_{a}F_{jb}
	+ F_{b}\,^{j}F_{aj} 
	+F^{j}\,_{b}F_{ja} 
	-\frac{1}{2}g_{ab}F_{ij}F^{ij})\big\}\, \sqrt{-g} \,\delta g^{ab}
\end{equation}
After simplifications, the energy-momentum tensor  takes the form
\begin{equation}
	T_{ab}=(3/8) \big\{\varphi^{5}F_{mn}F^{mn} (2F_{ak}F_{b}\,^{k}-\frac{1}{4}g_{ab}F_{ij}F^{ij})\big\}
\end{equation}

\par

$\bullet$   ${\cal I}_{4}$\\
Consider the variation of the fourth term as
\begin{equation}
	\begin{array}{ll}
	\delta {\cal I}_{4}= (5/8){\displaystyle \int }  d^{4}x \sqrt{-g}&\varphi^{5}(\delta F_{jm}F^{mn}F_{ni}F^{ij}+F_{jm}\delta F^{mn}F_{ni}F^{ij} \\ [4mm]
	&+F_{jm}F^{mn}\delta F_{ni}F^{ij}+F_{jm}F^{mn}F_{ni}\delta F^{ij})
		\end{array}
\end{equation}
To vary with respect to $A_{k}$, it can be observed that all of  the four terms above can be treated similarly.  Therefore it is sufficient to perform the variation only on one of  them.  For instance we may consider the third term, which  becomes
\begin{equation}
\delta {\cal I}_{43}	= (-5/4) \int  d^{4}x \, D_{i}(\sqrt{-g} \, \varphi^{5} F^{im}F_{mj}F^{jk})\,\delta A_{k}
\end{equation}
Treating the remaining terms in a similar way  and adding them up all,  this expression simplifies as
\begin{equation}
	a_{4}:=5 D_{j}(\sqrt{-g} \, \varphi^{5} F^{jm}F_{mi}F^{ik})
\end{equation}

\par  
Variation with respect to $\varphi$  yields
\begin{equation}
	\Phi_{4}:=(25/8)\varphi^{4} F_{jm}F^{mn}F_{ni}F^{ij}\sqrt{-g} 
\end{equation}	

We take the variation of  $\mathcal{L}_{4}$  with respect to metric $g^{ab}$,  to
obtain the energy-momentum tensor
\begin{equation}
		T_{ab}=(5/8)\varphi^{5}(2F_{am}F^{i}\,_{b}-\frac{1}{4}g_{ab}F_{jm}F^{ij}) \,F^{mn}F_{ni}
\end{equation}

\par


$\bullet$   ${\cal I}_{5}$ \\
The action for the fifth term is
\begin{equation}\label{i5}
	{\cal I}_{5}=\int d^{4}x\sqrt{-g} \,\varphi^{3}D_{k}F_{mn}D^{k}F^{mn}
\end{equation}
Lagrangians in the form of  $(DF)^{2}$ has been of interest  since long,  and is referred to as Bopp-Podolsky (BP) electrodynamics \cite{bopp1940, podolsky1942}.  
\par
Now , we vary (\ref{i5}) with respect to $A_{n}$ to  have
\begin{equation}
	{\delta\cal I}_{5}=4 \int d^{4}xD_{m}(\sqrt{-g} \, \varphi^{3}D_{k}D^{k}F^{mn})\, \delta A_{n}
\end{equation}
Thus, we find the expression
\begin{equation}
	a_{5}:=4D_{m}( \sqrt{-g} \varphi^{3} D_{k}D^{k}F^{mn})   
\end{equation}

\par
To vary  with respect to $\Gamma^{a}_{bc}$, we consider again
\begin{equation}
	{\delta\cal I}_{5}	=\int d^{4}x\sqrt{-g}\varphi^{3}[2\delta(D_{k}F_{mn})D^{k}F^{mn}]
\end{equation}
where
\begin{equation}
	\delta(D_{k}F_{mn})=-\delta\Gamma^{l}_{mk}F_{ln}-\delta\Gamma^{l}_{nk}F_{ml}
\end{equation}
After carrying out simple calculations, we obtain
\begin{equation}
	\gamma_{5}:=-4\sqrt{-g}\varphi^{3}F_{na}D^{c}F^{nb}
\end{equation}

There can be three distinguishable forms for $	\mathcal{L}$ in (\ref{i5}), in regard to their contracted indices, that are non-trivially equivalent
\begin{equation}
	\mathcal{L}^{(1)}=D_{k}F^{mk}D_{n}F_{m}\,^{n},\qquad
	\mathcal{L}^{(2)}=D_{k}F^{mn}D_{n}F_{m}\,^{k}, \qquad
	\mathcal{L}^{(3)}=D_{k}F^{mn}D^{k}F_{mn}
\end{equation}
We have $\mathcal{L}^{(3)}= 2 \mathcal{L}^{(2)}$.  The relation between $\mathcal{L}^{(1)}$ and $\mathcal{L}^{(2)}$ can be found by using total divergences, which becomes \cite{cuzinatto2018} :
\begin{equation}\label{difL}
	\mathcal{L}^{(2)}-\mathcal{L}^{(1)}=D_{n}(F_{m}\,^{k}D_{k}F^{mn})-D_{k}(F_{m}\,^{k}D_{n}F^{mn})+F_{m}\,^{k}(D_{k}D_{n}-D_{n}D_{k})F^{mn}
\end{equation}
By using
\begin{equation}
	(D_{k}D_{n}-D_{n}D_{k})F^{mn}=R^{m}\,_{lkn}F^{ln}-R^{ln}\,_{kn}F^{m}\,_{l}
\end{equation}\label{equivBP}
and arranging the indices we have an equivalent form for (\ref{i5})  
\begin{equation}\label{i5A}
	2	\int d^{4}x\sqrt{-g} \,\varphi^{3}  \{ R_{mn}F^{mj}F_{j}\,^{n}+R_{jmnk}F^{mk}F^{jn}+
	D_{n}F^{jn}D_{m}F_{j}\,^{m}\}
\end{equation} 
apart from total divergences, as usual.

To be on the safe side, we vary this  form with respect to $A_{k}$ and $\Gamma^{a}_{bc}$  in regard to the question of  whether an alternative form renders an expression comparable with the results of  \cite{baskal2013}. Since the results of our calculations are far from being affirmative, we do not write them here.

\par
From the variation of ${\cal I}_{5}$ with respect to $\varphi$, we get
\begin{equation}
	\Phi_{5}:=3\sqrt{-g}\varphi^{2} D_{k}F_{mn}D^{k}F^{mn}
\end{equation}

\par

Varying with respect to the metric the energy-momentum tensor takes the form
\begin{equation}
	T_{ab}=\varphi^{3}(\frac{1}{2}D_{a}F_{ij}D_{b}F^{ij}+D^{k}F_{a}\,^{j}D_{k}F_{bj}-\frac{1}{4}g_{ab}D^{k}F^{ij}D_{k}F_{ij})
\end{equation}

\par	


$\bullet $   ${\cal I}_{6}$ \\
For the sixth term with respect to $\varphi$,  we have
\begin{equation} \label{493}
	\delta {\cal I}_{6}=4\int d^{4}x \big\{-\varphi^{-2}(D_{m}D_{n}\varphi D^{m}D^{n}\varphi)\delta  \varphi +2\varphi^{-1}(D^{m}D^{n}\varphi)D_{m}D_{n}\delta  \varphi \big\} \sqrt{-g}
\end{equation}	
to obtain
\begin{equation}
	\Phi_{6}:=4\sqrt{-g}\big\{-\varphi^{-2}[(D_{m}D_{n}\varphi)D^{m}D^{n}\varphi]+2 D_{n}D_{m}(\varphi^{-1}D^{m}D^{n}\varphi)\big\}
\end{equation}

\par

Now varying with respect to $\Gamma^{a}_{bc}$
\begin{equation}
	\delta {\cal  I}_{6}=-8\int d^{4}x \,\sqrt{-g} \,\varphi^{-1}(\varphi_{a} D^{c}\varphi^{b})\delta \Gamma^{a}_{bc}
\end{equation}
we obtain
\begin{equation}
	\gamma_{6}:=-8\sqrt{-g}\varphi^{-1}(\varphi_{a} D^{c} \varphi^{b})
\end{equation}

We vary with respect to  $ g^{ab} $
\begin{equation}
\delta {\cal  I}_{6} =4\int d^{4}x  \varphi^{-1}
	\big\{D_{a}\varphi_{n}D_{b}\varphi^{n}+D_{m}\varphi_{a}D_{m}\varphi^{b}
	-\frac{1}{2}g_{ab}D_{m}\varphi_{n}D^{m}\varphi^{n}\big\}\sqrt{-g}\delta  g^{ab} 
\end{equation}
and use $D_{a}\varphi_{n}=D_{n}\varphi_{a}$  to obtain
\begin{equation}
	T_{ab}=4\varphi^{-1} (D_{a}\varphi_{n} D_{b}\varphi^{n}-\frac{1}{4}g_{ab}D_{m}\varphi_{n} D^{m}\varphi^{n})
\end{equation}

\par

$\bullet $  ${\cal I}_{7}$ \\
For the seventh term variation with respect to $A_{k}$ is
\begin{equation}
	\begin{array}{ll}
	\delta {\cal  I}_{7} = -2 {\displaystyle \int }  d^{4}x  \, \sqrt{-g} &
\varphi^{2} \, \big\{g^{mi}g^{ks}g^{np}[(D_{i}\delta A_{s}
	-D_{s}\delta A_{i})F_{pk} \\[3mm]
&	+(D_{p}\delta A_{k}-D_{k}\delta A_{p})F_{is}](D_{m}\varphi_{n})\big\}
	\end{array}
\end{equation}
yielding
\begin{equation}
	a_{7}:=-4 \sqrt{-g}D_{i}\big\{\varphi^{2} (F^{ki}D_{j}\varphi_{k}+F_{j}\,^{k}D^{i}\varphi_{k})\big\}
\end{equation}
From the variation 
with respect to $\Gamma^{a}_{bc}$,  we have the expression
\begin{equation}
	\gamma_{7}:=2\varphi^{2} F^{ck}F^{b}\,_{k}(D_{a}\varphi)\sqrt{-g}
\end{equation}

\par
Variation with respect to $\varphi$,
apart from the total divergences is
\begin{equation}
	{\delta\cal I}_{7}=-2 \int d^{4}x\sqrt{-g}\big\{D_{n}D_{m}(\varphi^{2} F^{mk}F^{n}\,_{k})+2\varphi F^{mk}F^{n}\,_{k}D_{m}\varphi_{n}\big\}\,\delta\varphi
\end{equation}
Then, we obtain
\begin{equation}
	\Phi_{7}:=-2 \sqrt{-g}\big\{D_{n}D_{m}(\varphi^{2} F^{mk}F^{n}\,_{k})+2 \varphi F^{mk}F^{n}\,_{k}D_{m}\varphi_{n} \big\}  
\end{equation}

\par

We consider $ \delta g^{ab}$ and proceed as we did earlier. So we  have
\begin{equation}
	T_{ab}=-\varphi^{2}\big\{F_{a}\,^{k}F^{i}\,_{k}D_{i}D_{b}\varphi
	+F_{b}\,^{k}F^{i}\,_{k}D_{i}D_{a}\varphi
	+F_{a}\,^{i}F_{b}\,^{j}D_{i}D_{j}\varphi
	-\frac{1}{2}g_{ab}F^{ik}F^{j}\,_{k}D_{i}D_{j}\varphi\big\}
\end{equation}

\par


$\bullet$   ${\cal I}_{8}$ \\
To our knowledge we have not encountered a Lagrangian containing such a term, apart from this particular KK reduction.  We start by varying with respect to the potentials $ A_{k}$. 
\begin{equation}
	\begin{array}{ll}
\delta	{\cal I}_{8}=4 {\displaystyle \int } d^{4}x\sqrt{-g} \,\varphi^{2} &\big\{ (D_{k}D_{m}\delta A_{n}-D_{k}D_{n}\delta A_{m})\varphi^{m}F^{kn}+D^{k}F_{m}\,^{n}\varphi^{m}(D_{k}\delta A_{n}-D_{n}\delta A_{k}) \\[4mm]
	& +	(D_{k}D_{m}\delta A_{n}-D_{k}D_{n}\delta A_{m})\varphi^{k}F^{mn}+D_{k}F^{mn}\varphi^{k}(D_{m}\delta A_{n}-D_{n}\delta A_{m})\big\}
	\\[2mm]
	\end{array}
\end{equation}
 
We consider
\begin{equation}\label{dtda3}
	\begin{array}{ll}
		D_{k}(\varphi^{2}\varphi^{m}F^{kn} D_{m}\delta A_{n})&=D_{m}\,[D_{k}(\varphi^{2}\varphi^{m}F^{kn})\, \delta A_{n}]
		-D_{m}D_{k}(\varphi^{2}\varphi^{m}F^{kn})\, \delta A_{n} \\[3mm]
		& + \varphi^{2}\varphi^{m}F^{kn} (D_{m}D_{k}\delta A_{n})
	\end{array}
\end{equation}

Therefore, we obtain 
\begin{eqnarray}
	\begin{array}{ll}
		a_{8}:=& 4  \Big\{D_{m}D_{k}[\varphi^{2}(F^{kn}\varphi^{m}+F^{mk}\varphi^{n}
		+2F^{mn}\varphi^{k} )]\\[3mm]
		&	+D_{k}[\varphi^{2}(D^{k}F^{n}\,_{m})\varphi^{m}
		+\varphi^{2}(D^{n}F_{m}\,^{k})\varphi^{m}] 
		+2D_{m}[\varphi^{2}(D_{k}F^{nm})\varphi^{k}]\Big\}\sqrt{-g}
	\end{array}
\end{eqnarray}

\par
In order to vary with respect to $\Gamma ^{a}_{bc}$ we rewrite $\delta	{\cal I}_{8}$  as
\begin{equation}
	\begin{array}{ll}
		\delta	{\cal I}_{8}&=4{\displaystyle \int } d^{4}x\sqrt{-g} \,\varphi^{2} \, \big\{(\partial_{k}F_{mn}-\delta\Gamma^{j}_{mk}F_{jn}-\delta\Gamma^{j}_{nk}F_{mj}[\varphi^{(m}F^{k)n}]\big\} \\[4mm]
		&=4{\displaystyle \int } d^{4}x\sqrt{-g} \varphi^{2}(-\delta ^{j}_{a}\delta ^{b}_{m}\delta ^{c}_{k}F_{jn}-\delta ^{j}_{a}\delta ^{b}_{n}\delta ^{c}_{k}F_{mj})[\varphi^{m}F^{kn}+\varphi^{k}F^{mn}]\delta \Gamma ^{a}_{bc}
	\end{array}				
\end{equation}
After simplifying, this becomes
\begin{equation}
	\gamma_{8}:=4\sqrt{-g}\varphi^{2}(\varphi^{b}F_{aj}F^{cj}-2\varphi^{c}F_{aj}F^{bj})
\end{equation}

\par
Variation with respect to $\varphi$  is
\begin{equation} 
	\delta	{\cal I}_{8}=4\int d^{4}x\sqrt{-g} \, (2 \varphi ^{-1}\delta \varphi L_{8}+\varphi^{2} D_{k}F^{m}\,_{n}D_{m}\delta \varphi F^{kn}+\varphi^{2} D^{k}F_{mn}D_{k}\delta \varphi F^{mn})
\end{equation}
Apart from total divergences, we have
\begin{equation}
	\delta  L_{8} =4\big\{(2 D_{k}F_{mn}\varphi^{(m} F^{k)n})-D_{m}[\varphi^{2}(D_{k}F_{mn})F^{kn}+\varphi^{2} (D^{m}F_{kn})F^{kn}]\big\}\delta \varphi  
\end{equation}
Then the expression becomes
\begin{equation}
	\Phi_{8}:=4\sqrt{-g}\big\{4 (D_{k}F_{mn}\varphi^{(m} F^{k)n})-D_{m}[\varphi^{2}(D_{k}F^{m}\,_{n})F^{kn}+\varphi^{2} (D^{m}F_{kn})F^{kn}]\big\}
\end{equation}

\par
We take the variation with respect to $g^{ab}$ to obtain
\begin{equation}
	\begin{array}{ll}
		T_{ab}&=2\varphi^{2}\big\{(D_{a}F^{mn}\varphi_{m}F_{bn}+D_{k}F_{an}\varphi_{b}F^{kn}+D_{k}F_{am}\varphi^{m}F_{b}\,^{k} \\ [2mm]
		&+D_{a}F_{mn}\varphi_{b}F^{mn}+2D_{k}F_{am}\varphi^{k}F_{b}\,^{m})-\frac{1}{2}g_{ab}D_{k}F_{mn}\varphi^{(m}F^{k)n}\big\}
	\end{array}
\end{equation}

\par

$\bullet$   ${\cal I}_{9}$  \\
We vary the ninth term with respect to $A_{n}$
\begin{equation}
\delta	{\cal I}_{9}=6 \int d^{4}x  \sqrt{-g} \, \big\{ 2 \varphi\varphi_{k}\varphi^{k}F^{mn}(D_{m}\delta A_{n})\big\}
\end{equation}
From above, we have the expression
\begin{equation}
	a_{9}:=-24\sqrt{-g}D_{m}(\varphi\varphi_{k}\varphi^{k}F^{mn})
\end{equation}

\par
Variation of ${\cal I}_{9}$ with respect to $\varphi$ yields
\begin{equation}
	{\delta\cal I}_{9}=6\int d^{4}x\big\{-2D_{k}(\sqrt{-g}\varphi F^{mn}F_{mn}\varphi^{k})+\varphi_{k}\varphi^{k}F^{mn}F_{mn}\sqrt{-g}\big\}\delta\varphi
\end{equation}
So, we have
\begin{equation}
	\Phi_{9}:=6 \big\{-2 D_{k}(\varphi F^{mn}F_{mn}\varphi^{k})+\varphi_{k}\varphi^{k}F^{mn}F_{mn}\big\}\sqrt{-g}
\end{equation}

\par
The energy-momentum tensor
\begin{equation}
	T_{ab}=6\varphi(\varphi_{k}\varphi^{k}F_{ai}F^{bi}
	+\frac{1}{2}\varphi_{a}\varphi_{b}F_{mn}F^{mn}
	-\frac{1}{4}g_{ab}\varphi_{k}\varphi^{k}F_{mn}F^{mn})
\end{equation}
is obtained by varying with respect to the metric, as before.\\

\par

$\bullet$    ${\cal I}_{10}$ \\
Varying  the last term ${\cal{I}}_{10}$  with respect to $A_{k}$ gives
\begin{equation}
	\begin{array}{ll}
		\delta	{\cal  I}_{10}=6{\displaystyle \int }  d^{4}x\sqrt{-g}
&\big\{-D_{m}(\varphi\varphi^{m}\varphi^{n}F_{n}\,^{k})
		+D_{m}(\varphi\varphi^{k}\varphi^{n}F_{n}\,^{m})\\[4mm]
		&-D_{n}(\varphi\varphi^{m}\varphi^{n}F_{m}\,^{k})
		+D_{n}(\varphi\varphi^{m}\varphi^{k}F_{m}\,^{n})\big\}  
		\sqrt{-g}\, \delta A_{k}  
	\end{array}
\end{equation}
Therefore, we have
\begin{equation}
	a_{10}:=12D_{m}(\varphi\varphi^{k}\varphi^{n}F_{n}\,^{m}-\varphi\varphi^{m}\varphi^{n}F_{n}\,^{k})\sqrt{-g}   
\end{equation}

Varying with respect to $\varphi$,  we obtain the expression
\begin{equation}
	\Phi_{10}:=6\Big\{\varphi_{m}\varphi_{n}F^{mk}F^{n}\,_{k}-	2D_{m}(\varphi\varphi_{n}F^{mk}F^{n}\,_{k})\Big\} \sqrt{-g}
\end{equation}

\par
Variation with respect to $g^{ab}$  
yields the energy-momentum tensor
\begin{equation}
	T_{ab}=6\varphi\Big\{\frac{1}{2}(\varphi_{i} \varphi_{j}F_{a}\,^{i}F_{b}\,^{j}+\varphi_{a}\varphi_{i}F_{bk}F^{ik} 
	+\varphi_{b}\varphi_{i}F_{ak}F^{ik})
	-\frac{1}{4}g_{ab}\varphi_{i}\varphi_{j}F^{ik}F^{j}\,_{k})\Big\}
\end{equation}

\section{The compatibility issue}

We have seen that in the standard KK theory, there are three sets of equations, irrespective of the order of the successive procedures.   They come from the variations with respect to three fields or from the splitting of $\hat G_{AB}$ into its two pure and one mixed components.   On the other hand, the usage of the Palatini method for the quadratic curvature model in \cite{baskal2013}, where we also consider the independent variation  of  the 5D connection $\hat{\Gamma}^{A}_{BC}$, yield six  equations.  Of  those three come from the splitting of  $D_{K}\hat R^{K}\,_{JMN}=0$, which are 
\begin{equation}
	D_{K}\hat R^{K}\,_{jmn}=0, \quad D_{K}\hat R^{K}\,_{5mn}=0 , \quad D_{K}\hat R^{K}\,_{5m5}=0
\end{equation}
since, from Bianchi identities  we have
\begin{equation} 
	D_{K}\hat R^{K}\,_{5mn}= 2 D_{K}\hat R^{K}\,_{[nm]5}
\end{equation}
reducing the number from four to three.
Another three  come  from (\ref{TABb}).   The component $T_{55}$ is included in (\ref{actionallterms}), apart from some numerical coupling constants.  A similar situation is valid for $G_{55}$ of (\ref{GAB}), and that it is included in (\ref{EHL}).  Obviously we do not expect an expression for
$T_{a5}$, coming from  this current procedure.

In order to compare our results with those in reference \cite{baskal2013}, we shall  simplify them by 
setting  the scalar field  $\varphi$  to be a constant.   This is the optimal amount of simplification,  for comparative purposes. 
Even in this simplified case, the non-minimal couplings between the gravitational field and the electromagnetic (EM) field are considerably intricate.   In addition to the pure gravitational part, it includes non-minimal couplings and non-linear electrodynamics.   Besides, we have the advantage of  ignoring the difference between the Einstein and Jordan frames.\\

 By taking account of  the calculations done for each term in the Lagrangian density  (\ref{actionallterms}), the field equation that comes from the variation of connection is
\begin{equation}\label{result1}
	D_{k}R^{k}\,_{jmn}=\varphi^{2} \left ( \frac{3}{4}D_{k}(F^{k}\,_{j}F_{mn}) - F^{k}\,_{[m}D_{n]}F_{kj}\right )
\end{equation}
This is the equation governing the gravitational filed. In \cite{pavelle1976}, Pavelle wrote:  {\sl "I will show that the KY equations require more restriction than simply the
elimination of degenerate spacetimes.  Indeed, it appears  that pure spaces themselves allow the generation of unphysical
solutions; and I suggest that one should not examine  $D_{k}R^{k}\,_{jmn}=S_{jmn}$  unless the source $S_{jmn}$ is nonvanishing." } Here, we see that the right hand side of 
(\ref{result1}) is the $S_{jmn}$ term Pavelle is emphasising,  whose existence and form appears naturally through the KK reduction procedure.

\par
On the other hand from $D_{K}\hat R^{K}\,_{jmn}=0 $, one has
\begin{equation}\label{BKone}
	D_{k}R^{k}\,_{jmn}=-\varphi^{2}  D_{[m}(F_{n]k}F^{k}\,_{j})
	-\frac{\varphi^{2}}{2} F_{j[m}D_{k}F^{k}\,_{|n]}
	+\frac{\varphi^{2}}{2}F_{mn}D_{k}F^{k}\,_{j}
\end{equation}
Equations (\ref{result1}) and (\ref{BKone}) do not seem to be equivalent, even if we consider equivalent expressions for the Lagrangian as in (\ref{i5A}).

As for  the equations governing the electromagnetic field, one may be tempted to take the variation directly with respect to $F^{mn}$ \cite{dereli2020}.   As it is clear in that article, in order to get the proper equations in regard to the electromagnetic equations, one should take an additional derivative.  If, such an approach  had been adopted then we would find 
\begin{equation}\label{same}
	D_{k}D^{k}F_{mn}= -\frac{3}{2}  R_{mnjk}F^{jk}+\frac{3}{4}  \varphi^{2} \, F_{mn} F^{jk}F_{jk}-
	\frac{5}{4} \varphi^{2} \, F_{mj}F^{jk}F_{kn}
\end{equation}
from the the variation with respect to $F^{mn}$ .  This   equation has the  same terms 
coming from the reduction of  $D_{K}\hat R^{K}\,_{5mn}=0 $, which explicitly reads as:
\begin{equation}
D_{k}D^{k}F_{mn}=- R_{mnjk}F^{jk}+\frac{\varphi^{2}}{2}(F_{mn}F_{jk}F^{jk}-2 F_{mj}F^{jk}F_{kn})
\end{equation}
but with different constants.   On the other hand, varying  with respect to $A_{n}$  renders
\begin{equation}
	D^{m}	D_{k}D^{k}F_{mn}= D^{m}\left\{ -\frac{3}{2} R_{mnjk}F^{jk}+\frac{3}{4} \varphi^{2} \,F_{mn}F^{jk}F_{jk} -
	\frac{5}{4} \varphi^{2} \, F_{mj}F^{jk}F_{kn} \right\}
\end{equation}
where, we observe that  (\ref{same})  just becomes its first integral.

\par

Similarly,  the overall energy-momentum tensor 
(again with  $ \varphi$  not depending on $x^{A}$) becomes
\begin{equation}	\label{tabfinal}
	\begin{array}{ll}
		T_{ab}&=\varphi \, \big \{(R_{anjk}R_{b}\,^{njk}-\frac{1}{4}g_{ab} R_{mnjk}R^{mnjk})\\ [3mm]
		&-\frac{3}{2} \varphi^{2}(R_{anjk}F_{b}\,^{n}+R_{bnjk}F_{a}\,^{n}-\frac{1}{4}g_{ab}R_{jkmn}F^{mn}) \, F^{jk}\\ [3mm]
		&+\frac{3}{8}\varphi^{4}(2F_{ak}F_{b}\,^{k}-\frac{1}{4}g_{ab}F_{ij}F^{ij}) F_{mn}F^{mn} \\[3mm]
		&	+\frac{5}{8}\varphi^{4}(2F_{am}F^{i}\,_{b}-\frac{1}{4}g_{ab}F_{jm}F^{ij}) \,F^{mn}F_{ni}\\ [3mm]
		&+\frac{1}{2}\varphi^{2}(D_{a}F_{ij}D_{b}F^{ij}+2D^{k}F_{a}\,^{j}D_{k}F_{bj}-\frac{1}{2}g_{ab}D^{k}F^{ij}D_{k}F_{ij})\big \}
	\end{array}
\end{equation}

One can immediately expect that when the Lagrangian density is reduced, there will be an additional $\varphi$ that multiplies all terms coming from the 4D form of the 5D $\sqrt{-\hat{g}}$.  Therefore, when the variation is taken with respect  to the 4D metric, there is an extra $\varphi$
appearing in $T_{ab}$, as we see in (\ref{tabfinal}) compared to 
$\hat T_{ab}$,  the reduced form of  $\hat T_{AB}$ to the actual spacetime.

\par
Now, let us consider the  spacetime components  of  (\ref{TABb})
\begin{equation}
	\hat{T}_{ab}=\hat{R}_{aKMN}\hat{R}_{b}\,^{KMN}
	-\frac{1}{4}\hat{g}_{ab}R_{JKMN}R^{JKMN}.
\end{equation}
The $T_{ab}$ component of  this expression is reduced to become \cite{baskal2013}:
\begin{equation}	
	\begin{array}{ll}
		^{R}T_{ab}&= \, \big \{(R_{anjk}R_{b}\,^{njk}-\frac{1}{4}g_{ab} R_{mnjk}R^{mnjk})\\ [3mm]
		&-\frac{3}{2} \varphi^{2}(\frac{1}{2}R_{anjk}F_{b}\,^{n}+\frac{1}{2}R_{bnjk}F_{a}\,^{n}-\frac{1}{4}g_{ab}R_{jkmn}F^{mn}) \, F^{jk}\\ [3mm]
		&+\frac{3}{8}\varphi^{4}(F_{ak}F_{b}\,^{k}-\frac{1}{4}g_{ab}F_{ij}F^{ij}) F_{mn}F^{mn}
		+\frac{5}{8}\varphi^{4}(F_{am}F^{i}\,_{b}-\frac{1}{4}g_{ab}F_{jm}F^{ij}) \,F^{mn}F_{ni}\\ [3mm]
		&+\frac{1}{2}\varphi^{2}(\frac{1}{2}D_{a}F_{ij}D_{b}F^{ij}+D^{k}F_{a}\,^{j}D_{k}F_{bj}-\frac{1}{2}g_{ab}D^{k}F^{ij}D_{k}F_{ij})\big \}
	\end{array}
\end{equation}
where  again we considered the case with $\varphi= constant $. 

We observe that, in our energy-momentum tensor,  (\ref{tabfinal}),  apart from the terms coming from the variation of $\sqrt{-g}$, there is a factor of  2  (preceding  some terms from lines two to last).   Except for that,  these two expressions are formally similar when compared with the results  given in \cite{baskal2013}.  The factor 2 comes from the fact that, when the action is reduced first, otherwise hidden $4D$ metrics come into play to form invariants from the constituent fields. \\

\section{Conclusion}

In this work, we have obtained the field equations and the energy-momentum tensor from the dimensionally reduced 5D WY action in order to compare its results with those obtained in \cite{baskal2013}.  Our current approach is to reverse the order of  variation and dimensional reduction mechanisms.  We see that both ways of ordering produce the same number of field equations, i.e., 3, however, now there is only one expression $T_{ab}$, while if  varied first, one has additional components $T_{a5}, T_{55}$.

 Although some striking similarities in the resulting field equations can be observed, it cannot be said that reversing the order of the applications yields the same equations.  This is basically due to the fact that when first reduced otherwise hidden fields of  the KK theory become emergent, as is seen in equation (\ref{actionallterms}),  and thus one has to take variations with respect to those hidden 4D metrics, connections  and potentials.  In addition to this, there are specific coupling constants for each term constituting the overall Lagrangian, that come into play and do not cancel out.  Since, the WY is quadratic, but not  linear as  the Einstein-Hilbert action,  it could be expected that some differences would occur.  However, one has to carry out the calculations in order to locate and describe the sources of  the differences and to obtain the exact forms of  the equations.

This work and  \cite{baskal2013}, use the Palatini method for variations. One may wish to examine  other variational methods for both ordering procedures, such as considering  an explicit dependence of  the connection on the metric, or introducing a torsion in the system.  Surely, the next natural step is to look for  non-trivial exact solutions for some specific metrics.

\end{document}